\documentclass[%
superscriptaddress,
preprintnumbers,
 amsmath,amssymb,
 aps,
pra,
]{revtex4-2}

\usepackage{graphicx}
\usepackage{dcolumn}
\usepackage{bm}
\usepackage{hyperref}
\usepackage[mathlines]{lineno}

\usepackage{siunitx}

\usepackage{comment}

\DeclareMathOperator\erf{erf}

\begin{document}

\preprint{APS/123-QED}

\title{Dissociative Recombination of Rotationally Cold ArH$^+$}

\author{\'Abel K\'alosi}
\email{abel.kalosi@outlook.com}
\affiliation{Columbia Astrophysics Laboratory, Columbia University, New York, NY 10027, USA}
\affiliation{Max-Planck-Institut f\"ur Kernphysik, Saupfercheckweg 1, D-69117 Heidelberg, Germany}

\author{Manfred Grieser}
\affiliation{Max-Planck-Institut f\"ur Kernphysik, Saupfercheckweg 1, D-69117 Heidelberg, Germany}

\author{Leonard W. Isberner}
\affiliation{I. Physikalisches Institut, Justus-Liebig-Universit\"at Gie{\ss}en, D-35392 Gie{\ss}en, Germany}
\affiliation{Max-Planck-Institut f\"ur Kernphysik, Saupfercheckweg 1, D-69117 Heidelberg, Germany}

\author{Holger Kreckel}
\affiliation{Max-Planck-Institut f\"ur Kernphysik, Saupfercheckweg 1, D-69117 Heidelberg, Germany}

\author{\r{A}sa Larson}
\affiliation{Department of Physics, Stockholm University, AlbaNova University Center, SE-106 91 Stockholm, Sweden}

\author{David A. Neufeld}
\affiliation{Department of Physics \& Astronomy, Johns Hopkins University, Baltimore, MD 21218, USA}

\author{Ann E. Orel}
\affiliation{Department of Chemical Engineering, University of California, Davis, CA 95616, USA}

\author{Daniel Paul}
\affiliation{Columbia Astrophysics Laboratory, Columbia University, New York, NY 10027, USA}
\affiliation{Max-Planck-Institut f\"ur Kernphysik, Saupfercheckweg 1, D-69117 Heidelberg, Germany}

\author{Daniel W. Savin}
\affiliation{Columbia Astrophysics Laboratory, Columbia University, New York, NY 10027, USA}

\author{Stefan Schippers}
\affiliation{I. Physikalisches Institut, Justus-Liebig-Universit\"at Gie{\ss}en, D-35392 Gie{\ss}en, Germany}

\author{Viviane C. Schmidt}
\affiliation{Max-Planck-Institut f\"ur Kernphysik, Saupfercheckweg 1, D-69117 Heidelberg, Germany}

\author{Andreas Wolf}
\affiliation{Max-Planck-Institut f\"ur Kernphysik, Saupfercheckweg 1, D-69117 Heidelberg, Germany}

\author{Mark G. Wolfire}
\affiliation{Department of Astronomy, University of Maryland, College Park, MD 20742, USA}

\author{Old{\v r}ich Novotn\'y}
\affiliation{Max-Planck-Institut f\"ur Kernphysik, Saupfercheckweg 1, D-69117 Heidelberg, Germany}

\date{\today}

\begin{abstract}

We have experimentally studied dissociative recombination (DR) of electronically and vibrationally relaxed ArH$^+$ in its lowest rotational levels, using an electron--ion merged-beams setup at the Cryogenic Storage Ring. We report measurements for the merged-beams rate coefficient of ArH$^+$ and compare it to published experimental and theoretical results. In addition, by measuring the kinetic energy released to the DR fragments, we have determined the internal state of the DR products after dissociation. At low collision energies, we find that the atomic products are in their respective ground states, which are only accessible via non-adiabatic couplings to neutral Rydberg states. Published theoretical results for ArH$^+$ have not included this DR pathway. From our measurements, we have also derived a kinetic temperature rate coefficient for use in astrochemical models.

\end{abstract}

\maketitle


\section{\label{sec:intro} Introduction}

\begin{figure}[ht!]
\includegraphics{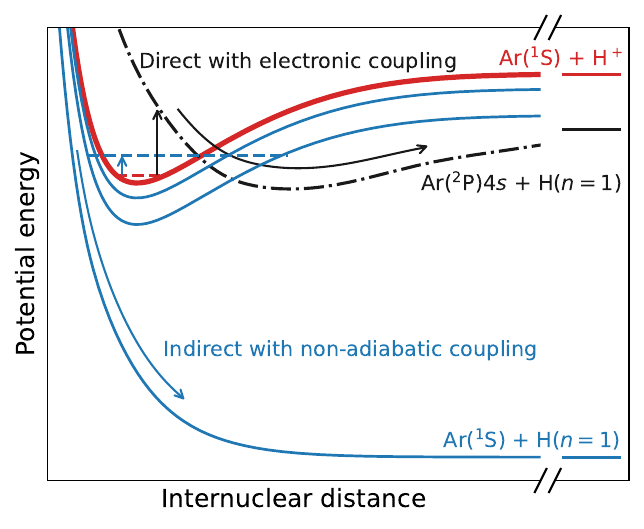}
\caption{Schematic PECs illustrating \textbf{some of the possible} DR mechanisms of ArH$^+$ at low and high collision energies, shown by the short blue and long black arrow pairs, respectively. The thick red curve is the PEC of the target ion. Below the ionic PEC are Rydberg PECs of the neutral system, which have similar shapes to the ionic PEC. For clarity, only two of the infinte number of Rydberg PECs are shown by thin blue curves. The horizontal dashed lines highlight selected ro-vibrational levels of the ionic and Rydberg PECs to which they are connected. At low collision energies, the electron is captured into a ro-vibrationally excited Rydberg state. The system then pre-dissociates through non-adiabatic couplings to the ground state (left-side blue pathway). At higher collision energies, the electron can be captured into an electronic resonant state, shown by the dot-dashed black curve, that leads directly to dissociation (right-side black pathway). On the right-hand side of the figure, the horizontal full lines show the dissociation limits for the neutral and ionic ground states and the electronic resonant state, shown by the blue, red, and black, respectively.
\label{fig:schempot}}
\end{figure}

Dissociative recombination (DR) with free electrons is an important neutralization process for molecular cations in a variety of laboratory, industrial, and astrophysical environments \cite{geppert_drinter_2008}.
Historically, DR was proposed to proceed via electron capture into a neutral dissociative (resonance) state with a potential energy curve (PEC) that crosses the PEC of the target ion. It was soon recognized that the incoming electron could also be captured into a ro-vibrationally excited Rydberg state, forming an intermediate step that then pre-dissociates via a suitable state. These two pathways were dubbed the direct and indirect DR mechanisms \cite{florescumitchell_dr_2006, drbook} and are illustrated schematically in Fig.~\ref{fig:schempot}. Both mechanisms occur simultaneously and interference between them will cause sharp structures observed in high resolution  cross section measurements. Since the two mechanisms cannot be separated, modern theories strive \textbf{either to improve the existing unified treatments or to elaborate new ones that are more generally applicable or accurate (e.g., Ref~\cite{forer_unified_2023} and references therein)}.

In some systems, though, direct DR is significantly suppressed. Either there is no electronic resonant state with a PEC crossing the ionic PEC, or the resonant state PEC does not cross the ionic PEC close to its equilibrium internuclear distance, the latter of which is illustrated schematically in Fig.~\ref{fig:schempot}. Instead, electron capture at low energies is driven by recombination into a ro-vibrationally excited Rydberg state of the neutral molecule. This Rydberg state pre-dissociates by non-adiabatic couplings to a state that is open for dissociation \cite{guberman_dr_1994}. This mechanism is illustrated with the blue arrows in Fig.~\ref{fig:schempot} and occurs for certain rare gas diatomic hydrides \cite{mitchell_dissneh_2005, mitchell_diss_2005, curik_dissociative_2020}. Direct DR is negligible in these systems at low collision energies. To theoretically model the indirect DR process, reliable descriptions are needed for the non-adiabatic couplings among the series of infinite numbers of bound Rydberg states, as well as the coupling to the ionization continuum. Recent multi-channel quantum defect theory (MQDT) calculations on the relatively simple HeH$^+$ system \cite{curik_dissociative_2020} have highlighted the importance of using an optimal origin for the coordinate system in which the quantum defects of the Rydberg states are calculated. These calculations also predicted a considerable influence of the rotational structure on the DR cross section. This had already been observed experimentally in rotational-level-resolved DR measurements for HeH$^+$ that were enabled by recent laboratory advances \cite{novotny_quantum_2019}. Good agreement between theory and experiment for this simple system was achieved for cross sections at collision energies as low as a few meV \cite{curik_dissociative_2020}.

Here, we move to a theoretically more challenging system  with a higher level of electronic complexity and report laboratory measurements for DR of ground state ArH$^+$, which can be schematically represented as
\begin{equation}\label{eq:dr}
    \mathrm{ArH}^+ + \mathrm{e}^- \rightarrow \mathrm{Ar} + \mathrm{H}.
\end{equation}
The only published experimental data are the results of a merged-beams experiment from the room-temperature ASTRID storage ring \cite{mitchell_diss_2005}. However, their measurements could only put an upper limit on the strength of the DR process at collision energies $\lesssim 1$~eV, where the indirect process is expected to dominate in ArH$^+$.

To the best of our knowledge, no published theoretical calculations are available for low energy DR of ArH$^+$. Recent theoretical MQDT calculations included dissociation through several resonance states of ArH that cross the ionic ground state PEC at internuclear distances much larger than the equilibrium distance of ArH$^+$ \cite{abdoulanziz_theor_2018, Djuissi_2022}. Given that the resonance states considered correlate with dissociation limits located above the ionic ground vibrational level, the calculated DR cross section is only non-zero when these channels become energetically open ($\gtrsim 1$~eV) \cite{abdoulanziz_theor_2018, Djuissi_2022}.

Renewed interest in DR of ArH$^+$ is also linked to its astrophysical application. $^{36}$ArH$^+$ was the first noble gas molecule to be detected in interstellar space \cite{barlow_detection_2013}, where it was seen in emission. Subsequently, two isotopologues, $^{36}$ArH$^+$ and $^{38}$ArH$^+$, were observed in absorption in astronomical spectra \cite{schilke_argonium_2014, jacob_extending_2020}, showing that ArH$^+$ can be produced in a variety of environments, ranging from diffuse clouds to supernova remnants. The relatively low DR rate coefficient of ArH$^+$ is expected to contribute to its survival at conditions where most other diatomic species would be destroyed at rates more than two orders of magnitude faster. Recent astrochemical models have adopted the upper limit of the room temperature ASTRID measurements over a wide range of temperatures \cite{neufeld_cheminterarg_2016, priestley_modelling_2017}, leading to two potential complications. In interstellar clouds, where collisional timescales are much longer than radiative lifetimes, only the lowest rotational level of ArH$^+$ is populated, and state-specific data are needed. In environments with kinetic temperatures of $\sim 1000$~K, a reliable temperature dependence is needed. In both instances, the ASTRID results do not accommodate the needs of the astronomical community.

Our work here addresses both the molecular physics and astrophysics issues related to DR of ArH$^+$. The remainder of this paper is structured as follows. In Sec.~\ref{sec:exp}, we briefly discuss the experimental setup and data analysis. In Sec.~\ref{sec:res}, we present the measured merged-beams rate coefficient, fragment-imaging results, and the derived kinetic temperature rate coefficient for plasma applications. In Sec.~\ref{sec:pec}, calculated potential energy curves of the ArH and ArH$^+$ systems relevant for the low energy DR are presented. 
In Sec.~\ref{sec:discuss}, the present results are discussed in the context of DR theory and astrochemistry. A summary is given in Sec.~\ref{sec:summ}.

\section{\label{sec:exp} Experimental Description}

DR measurements were performed on internally cold $^{40}$ArH$^+$ using the Cryogenic Storage Ring (CSR) \cite{vonhahn_csr_2016} facility at the Max Planck Institute for Nuclear Physics in Heidelberg, Germany. We followed the methodology of our earlier measurements to determine the merged-beams rate coefficient and to perform imaging of the DR products \cite{novotny_quantum_2019, paul_dr_2022, jain_tio_2023}. Here we discuss only those new aspects specific to the present measurements.

\subsection{\label{subsec:ion} Ion Beam and Storage}

$^{40}$ArH$^+$ ions were generated in an electron cyclotron resonance (ECR) ion source\textbf{, using argon buffer gas with a few percent admixture of hydrogen}. A current of up to a few hundred nA was extracted, accelerated to an energy of $\approx 300$~keV, mass-to-charge selected using a series of dipole magnets, and then injected into CSR. We typically injected around $1\times10^7$ ions, in order to remain within the linear counting regime of our neutral particle detector. The injected ions were stored on a closed orbit in CSR and merged with a magnetically confined electron beam in the electron cooler straight section of the ring. The storage ring lattice was set up in the so-called achromat mode, a configuration where the momentum dispersion was near zero in the electron--ion overlap section. This mode was necessary for efficient phase-space cooling of the ArH$^+$ ion beam, also know\textbf{n} as electron cooling \cite{poth_cooling_1990}. Electron cooling enables us to reduce the diameter and energy spread of the stored ion beam. We typically observed exponential beam lifetimes of $> 600$~s for electron-cooled beams. For data collection, we stored the ions for times of up to $450$~s.

Gas-discharge ion sources produce internally excited ions. Once the ions were stored, they rapidly relaxed to their ground electronic and vibrational states, within a few ms after injection. Rotational relaxation was driven by spontaneous emission in the cryogenic black-body environment of CSR \cite{oconnor_photo_2016, meyer_radiative_2017, kalosi_inelastic_2022} on timescales of tens of seconds. Additionally, the stored ions benefited from rotational cooling through electron--ion rotational-level-changing collisions using the electron beam. A much colder rotational distribution was reached than that of the ASTRID ArH$^+$ study \cite{mitchell_diss_2005}. Using our collisional-radiative model for the populations of rotational levels \cite{kalosi_inelastic_2022}, labelled by quantum number $J$ for ArH$^+$, we estimate that $\approx 57\%$ of the ions were in the $J=0$ ground level, $\approx 37\%$ in the $J=1$ level, and $\approx 6\%$ in the $J=2$ level during data collection for DR (see Appendix~\ref{appsec:rotmodel}).

We used a high purity beam of $^{40}$ArH$^+$ that was monitored in situ during ion storage. The initial beam purity was verified by operating CSR in the recently developed isochronous mode, i.e., as a time-of-flight mass spectrometer with a mass resolution of $\sim 10^{-5}$~u \cite{grieser_iso_2022}. A particular concern was contamination by potential impurities in the ion source discharge, namely $^{12}$C$_3$H$_5^+$ and $^{12}$CH$_3$CN$^+$, which differ in mass from $^{40}$ArH$^+$ by $\approx 7\times10^{-2}$~u and $\approx 6\times10^{-2}$~u, respectively. To enhance the purity of the stored beam, we used the electron beam to deplete these impurities, which have orders of magnitude higher DR rate coefficients compared to ArH$^+$. Using the imaging techniques introduced in Appendix~\ref{appsec:img}, a detailed contaminant analysis is given in Appendix~\ref{appsec:contam}.

\subsection{\label{subsec:mbexp} Merged-beams Experiment}

DR measurements were performed using an electron--ion merged-beams configuration that has been discussed in detail elsewhere \cite{novotny_quantum_2019, paul_dr_2022, jain_tio_2023}. In the present measurements, the relative collision energy between the merged beams was controlled by varying the voltage on a set of biased drift tubes in the electron cooler. This enabled us to tune the nominal laboratory-frame electron-beam energy in the interaction region, $E_\mathrm{e}$. The drift tubes define the effective interaction region of the merged-beams setup. DR reactions produced neutral fragments, see Reaction~(\ref{eq:dr}), that retained nearly the same laboratory-frame velocity as the reacting ion and continued on ballistic trajectories. The neutral reaction products were collected by a particle-counting and impact-position-imaging detector that was located $3.8$~m downstream of the electron--ion interaction region and oriented to face perpendicular to the beam direction. By recording the count rate of DR reactions as a function of the relative electron--ion collision velocity $v_\mathrm{r}$, we measured the merged-beams rate coefficient
\begin{equation}
    \alpha^\mathrm{mb} = \langle \sigma v_\mathrm{r} \rangle,
\end{equation}
where the product of $v_\mathrm{r}$ and the energy dependent DR cross section $\sigma$ is averaged over the velocity distribution in the electron--ion overlap section. The main experimental factors determining the velocity distribution are the perpendicular and parallel temperature components $T_\perp$ and $T_\parallel$, respectively, in the frame of the bulk electron-velocity vector; the merging and demerging geometry; and the variable laboratory-frame energy of the electrons, $E_\mathrm{e}$. We modelled the velocity distribution following our previous approaches \cite{novotny_drhcl_2013, kalosi_inelastic_2022}. Additionally, it is practical to express the collision-energy distribution as a function of the detuning energy $E_\mathrm{d}$, defined as the nominal center-of-mass collision energy, 
\begin{equation}\label{eq:det}   
    E_\mathrm{d} = \left(\sqrt{E_\mathrm{e}} - \sqrt{E_0} \right)^2,
\end{equation}
where $E_0$ is the laboratory-frame electron-beam energy at matched electron--ion velocities, substituting the reduced mass of the collision system by that of the electron.

For the present measurements, we used a magnetically confined electron beam with a current of $9.15$~{\textmu}A and a circular density profile that was nearly uniform. The initial laboratory-frame energy of the electrons, $\approx 7$~eV, was chosen so that the merging-geometry-dependent collision energies outside the effective interaction region were $\lesssim 0.5$~eV, i.e., below the first excited DR product threshold (see Table~\ref{tab:drenergy}). The electrons were guided into CSR by a step-wise reduction in magnetic field strength, decreasing to $10$~mT in the interaction region. The total adiabatic expansion factor was $30$, resulting in an expanded beam diameter of $12.1\pm0.6$~mm. For the perpendicular electron-beam temperature, we take $k_\mathrm{B}T_\perp  = 2.0\pm1.0$~meV \cite{kalosi_diss_2023}. The parallel electron-beam temperature is energy dependent \cite{paul_thesis_2021}, with a value of $k_\mathrm{B}T_\parallel \approx 0.26$~meV at matched velocities.

\begin{table}[b]
\renewcommand*{\arraystretch}{1.4}
\caption{\label{tab:drenergy}%
Reaction threshold energies $E_\mathrm{r}$ according to Eq.~(\ref{eq:re}) for the four lowest excited state products for DR of ground state ArH$^+$. The reaction energies for Ar excited states are calculated as the weighted average of the $J$ sublevel energies.
}
\begin{ruledtabular}
\begin{tabular}{ccc}
\textrm{Ar}&
\textrm{H}&
$E_\mathrm{r}$ (eV)\\
\colrule
$\left[\mathrm{Ne}\right]3s^23p^6(^1\mathrm{S})$ & $n=2$ & 0.5 \\
$\left[\mathrm{Ne}\right]3s^23p^5(^2\mathrm{P}^\circ_{3/2})4s$ & $n=1$ & 1.9 \\
$\left[\mathrm{Ne}\right]3s^23p^5(^2\mathrm{P}^\circ_{1/2})4s$ & $n=1$ & 2.1 \\
$\left[\mathrm{Ne}\right]3s^23p^6(^1\mathrm{S})$ & $n=3$ & 2.4 \\
\end{tabular}
\end{ruledtabular}
\end{table}

Electron cooling occurs for matched electron-ion velocities and was utilized for all DR measurements. Using Schottky pickup observations \cite{vonhahn_csr_2016}, we measured the revolution frequency and momentum spread of the stored ions. The corresponding laboratory-frame electron-beam energy was $E_0 = 4.003\pm0.012$~eV at matched velocities, also called the cooling energy. For the DR measurements, we applied uninterrupted electron cooling immediately after ion injection, for a duration of $42$~s. This resulted in the reduction of the injected ion-beam horizontal and vertical full width at half maxima (FWHM) to $<5.2\pm0.7$~mm and $<2.8\pm0.5$~mm, respectively.

Measurements of the collision-energy-dependent merged-beams rate coefficient were performed as a function of storage time, starting at $42$~s after ion injection, by periodically detuning the electron-beam energy $E_\mathrm{e}$ from matched velocities. The measurements were grouped into predefined sets of $E_\mathrm{d}$ values. Data were collected for $25$~ms at a given measurement $E_\mathrm{d}$, followed by electron cooling for $100$~ms, a reference energy of $E_\mathrm{d} = 30$~eV for $25$~ms, and electrons off for $25$~ms. We also included an $\sim 5$~ms waiting time between each of these steps for the system to stabilize.

The measured absolute merged-beams rate coefficient,
\begin{equation}\label{eq:mbrc}
    \alpha^\mathrm{mb}(E_\mathrm{d}) = \frac{R_\mathrm{e}(E_\mathrm{d})}{\eta(E_\mathrm{d}) \xi N_\mathrm{i} n_\mathrm{e}(E_\mathrm{d}) \hat{l}_0/C_0},
\end{equation}
is directly proportional to the electron-induced count rate $R_\mathrm{e}(E_\mathrm{d})$. At most energies $R_\mathrm{e}(E_\mathrm{d})$ is due to DR. Above $3.9$~eV, it may be partly also due to dissociative excitation (DE). Quantities that are energy dependent, such as the detection efficiency for DR events $\eta(E_\mathrm{d})$ and the electron density $n_\mathrm{e}(E_\mathrm{d})$, are discussed below. The fraction of ions enclosed by the electron beam in the effective interaction region, $\xi > 0.992$, was determined from the profiles of the ion and electron beams. $N_\mathrm{i}$ is the number of stored ions. The effective electron--ion overlap length is measurement specific; here $\hat{l}_0 = 0.79\pm0.01$~m. It has been previously defined in Ref.~\cite{kalosi_inelastic_2022}. The circumference of the ion-beam orbit is $C_0 = 35.12\pm0.05$~m.

Additional background processes, due to residual-gas-induced collisions and the intrinsic dark rate of the neutral detector, contribute to the measured count rate during a DR measurement step. $R_\mathrm{e}(E_\mathrm{d})$ was determined after subtracting the count rate due to background contributions measured in the corresponding electrons-off step.

We determined $\eta(E_\mathrm{d})$ by comparing single and double particle hits on the detector \cite{paul_dr_2022}. Due to the finite size of the particle counting detector, $\eta(E_\mathrm{d})$ is dependent on the kinetic energy released (KER) in the DR process and the branching ratios between dissociation channels. For additional details see Appendix~\ref{appsec:img}. At matched velocities, we obtained $\eta(0~\mathrm{eV}) = 0.62$ and applied this value for all $E_\mathrm{d} < 1.85$~eV. Above this energy, the $\mathrm{Ar}(^2\mathrm{P})4\mathrm{s} + \mathrm{H}(n=1)$ DR channels open and dominate the measured particle flux (see Sec.~\ref{subsec:imgdata}). We analyzed $\eta(E_\mathrm{d})$ at selected energies from $3$~eV to $7.5$~eV and found values that decreased monotonically from $0.87$ to $0.65$. We constructed a geometrical detector-cutoff model, assuming an isotropic transverse fragment distribution \cite{jain_tio_2023} and taking into account the increase of the KER with $E_\mathrm{d}$, to find a smooth representation for $\eta(E_\mathrm{d})$ from $1.85$~eV to $7$~eV. Above $7$~eV, we used a single effective value of $0.64$.

Equation~(\ref{eq:mbrc}) was initially evaluated on a relative scale by replacing the storage-time-dependent ion number $N_\mathrm{i}(t)$ with an electron-induced neutralization signal. In specific, we normalized using the count rate at the reference energy of $E_\mathrm{d} = 30$~eV, $R_\mathrm{ref}(t)$, determined from the reference rate minus the electrons-off rate. $R_\mathrm{ref}$ was measured at a sufficiently high energy that the neutralization rate coefficient can be assumed to be independent of the internal state of the ions. The $\alpha^\mathrm{mb}(E_\mathrm{d})$ measurements were accompanied by a calibration of $R_\mathrm{ref}(t)$ to $N_\mathrm{i}(t)$, expressed as a proportionality factor $S_\mathrm{b} = R_\mathrm{ref}(t)/N_\mathrm{i}(t)$. $N_\mathrm{i}$ was measured using beam bunching combined with a capacitive current pickup (PU-C in Ref.~\cite{vonhahn_csr_2016}). The absolute current measurement for PU-C was calibrated with a $10\%$ systematic uncertainty \cite{paul_dr_2022}. The determined $S_\mathrm{b}$ is specific to the present ArH$^+$ campaign as it is proportional to the electron density and the neutralization rate coefficient for ArH$^+$ at $E_\mathrm{d} = 30$~eV. To verify the linear behaviour of $S_\mathrm{b}$, we used ion-beam bunching at various storage times and for ion numbers in the range of $3 \times 10^6$ to $1.2 \times 10^7$. The uncertainty due to beam bunching introduced an additional $15\%$ uncertainty when evaluating $S_\mathrm{b}$.

The $\alpha^\mathrm{mb}(E_\mathrm{d})$ measurements were performed using a constant electron current. Combined with the measured electron-beam radius, we determined the energy dependent $n_\mathrm{e}(E_\mathrm{d})$ by taking into account the laboratory-frame electron energy versus $E_\mathrm{d}$. The measurement value at matched velocities was $n_\mathrm{e}(0~\mathrm{eV}) = (4.2\pm0.4)\times10^{5}$ cm$^{-3}$.

The total systematic uncertainty of the absolute scaling for our merged-beams DR rate coefficient $\alpha^\mathrm{mb}$ was determined by adding all contributing uncertainties listed above in quadrature. The resulting total systematic uncertainty of $21\%$ is valid for $E_\mathrm{d} < 7$~eV.

The particle-counting detector used in the present study is also capable of imaging the impact positions and measuring the temporal separation of the neutral DR fragments. The 120~mm diameter microchannel plate (MCP) detector was backed by a phosphor anode that converted the electron cascade signal from the MCP into light spots, signifying the impact positions, that were observed with an optical camera. The particle detection electronics were triggered by a silicon photomultiplier that viewed the phosphor screen and was used to synchronize the optical camera and electronic readout systems. Impact positions were read out by a 1 kHz frame-rate optical camera, with an exposure time of $\approx 2$~{\textmu}s. Arrival times were determined by analyzing the MCP electronics signal. Additional details about the imaging setup can be found in Ref.~\cite{paul_thesis_2021}. For DR, the typical temporal separation between the Ar and H fragments was $< 150$~ns. By measuring the distance and time between the fragments, we generated three-dimensional (3D) images of the dissociating fragments. From these data we determined the KER in the DR process and identified the internal state of the products after dissociation.
   
\section{\label{sec:res} Experimental Results}

\subsection{\label{subsec:mbdata} Merged-beams Rate Coefficient}

\begin{figure*}[ht!]
\includegraphics{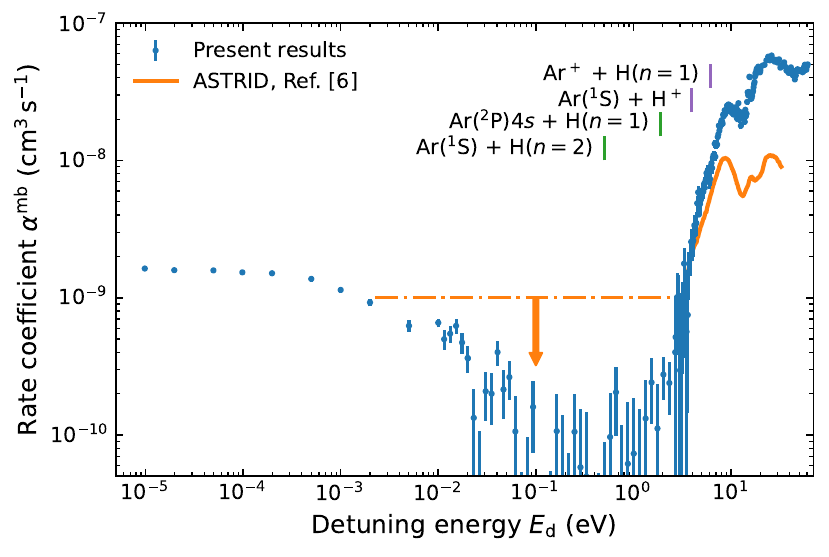}
\caption{Measured merged-beams rate coefficient $\alpha^\mathrm{mb}$ for neutralization of ArH$^+$ ions in collisions with electrons vs. detuning energy $E_\mathrm{d}$. The present results are plotted as blue symbols with error bars representing one-sigma statistical uncertainties. The absolute scaling of the results has a systematic accuracy of $21\%$ at $E_\mathrm{d} \leq 7$~eV. Also shown are the results from the room-temperature ASTRID study, indicated by the full orange line \cite{mitchell_diss_2005}. The ASTRID data are plotted only for values above $1\times10^{-9}$~cm$^{3}$~s$^{-1}$, the corresponding low energy results are represented by the horizontal dot-dashed line. The vertical lines with labels highlight reaction threshold energies $E_\mathrm{r}$ for selected DR and DE products. \textbf{Our measured $\alpha^\mathrm{mb}$ is provided in tabulated form in Ref.~\cite{supplement}.}
\label{fig:mbrcA}}
\end{figure*}

The measured merged-beams rate coefficient $\alpha^\mathrm{mb}(E_\mathrm{d})$ for reactive collisions of ArH$^+$ with electrons is plotted in Fig.~\ref{fig:mbrcA}. We first compare our results with those measured at the room-temperature ASTRID storage ring \cite{mitchell_diss_2005}. At the lowest resolved energies ($\sim 10^{-3}$~eV), no measurable signal was detected in the ASTRID measurements and an upper limit of $1\times10^{-9}$~cm$^{3}$~s$^{-1}$ was determined \cite{mitchell_diss_2005}. In our measurements, $\alpha^\mathrm{mb}$ is approximately constant below $10^{-3}$~eV, as expected for our energy resolution. Between $0.1$ and $1$~eV, the CSR values are almost an order of magnitude lower than the low-energy limit of the ASTRID data. Here, our statistical counting uncertainties vary from $0.9\times10^{-10}$ to $1.1\times10^{-10}$~cm$^{3}$~s$^{-1}$. By calculating an energy independent mean from the CSR DR signal rate between $0.1$ and $1$~eV, we derive a limit of $(0.6\pm2.4)\times10^{-11}$~cm$^{3}$~s$^{-1}$, which is a more than an order of magnitude improvement over the ASTRID limit.

At energies $>3.9$~eV, DE has to be considered as an additional process that results in neutral fragments and can contribute to the measured signals. The CSR particle detection system is currently not capable of distinguishing between DR and DE, while the ASTRID experiments used a mass-sensitive detection method that could distinguish between DR and those DE events that result in the production of only neutral H \cite{mitchell_diss_2005}. Still, both measurements potentially included the DE channel
\begin{equation}\label{eq:deAr}
    \mathrm{ArH}^+ + \mathrm{e}^- \rightarrow
    \mathrm{Ar} + \mathrm{H}^+ + \mathrm{e}^-,
\end{equation}
which is energetically accessible for collision energies above $3.9$~eV.

Between $\approx 3$ and 6~eV, we find reasonable agreement between the shape of the CSR and ASTRID DR results, taking into account the uncertainty in the absolute scaling of the ASTRID measurements. Above $\approx 6$~eV, the CSR and ASTRID results gradually diverge. We attribute this to the opening of the DE channel
\begin{equation}\label{eq:deH}
    \mathrm{ArH}^+ + \mathrm{e}^- \rightarrow
    \mathrm{Ar}^+ + \mathrm{H} + \mathrm{e}^-,
\end{equation}
which has a threshold of $6.1$~eV and could be discriminated against in the ASTRID results. 

\begin{figure}[ht!]
\includegraphics{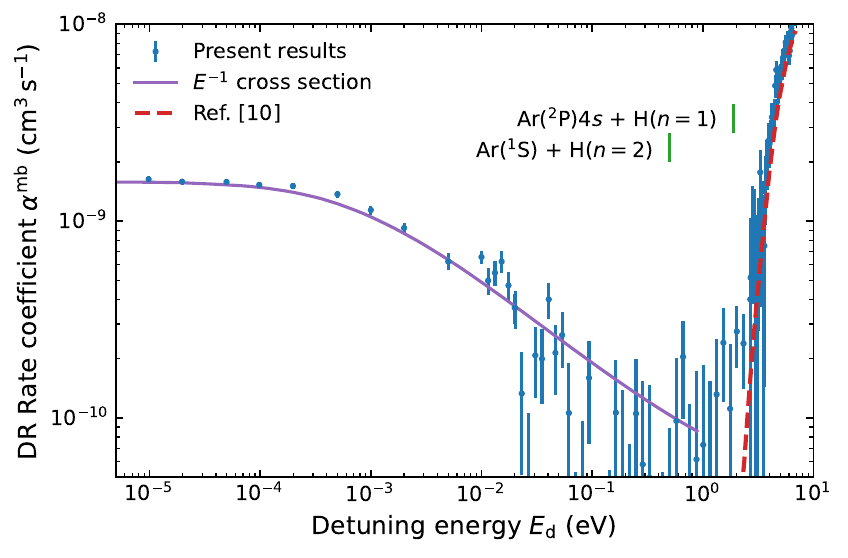}
\caption{Same as Fig.~\ref{fig:mbrcA} but comparing with two different theoretical cross section models. The theoretical results have been transformed into a merged-beams rate coefficient using our collision velocity distribution.
\label{fig:mbrc}}
\end{figure}

A comparison of the measured DR rate coefficient $\alpha^\mathrm{mb}(E_\mathrm{d})$ with two different models is plotted in Fig.~\ref{fig:mbrc}. The cross sections were converted into $\alpha^\mathrm{mb}$ using our collision velocity distribution. For $E_\mathrm{d} < 1$~eV we used a structureless cross section model given by a collision energy dependence of $E^{-1}$ and scaled to the measured data at matched velocities. This model represents electron capture without taking into account the internal structure of the target ion or any relevant intermediate states \cite{drbook}. For $E_\mathrm{d} \gtrsim 2$~eV, we can compare to the theoretical calculations of Refs.~\cite{abdoulanziz_theor_2018, Djuissi_2022}, which show good agreement with our experimental results, to within our systematic uncertainty of $21\%$. This suggests that the DE channels of Eqns.~(\ref{eq:deH}) and~(\ref{eq:deAr}) contribute at most at the level of the systematic uncertainty in the measured $\alpha^\mathrm{mb}$ for $E_\mathrm{d}$ between $3.9$ and $6.0$~eV.

\subsection{\label{subsec:imgdata} 3D Imaging}

We determined the internal state of the DR products using the energy $E_\mathrm{KER}$ of the DR fragments. For this, we acquired 3D imaging data from the measured impact distances and temporal separations. Dedicated measurements were performed for several values of $E_\mathrm{d}$. To account for the finite electron--ion overlap length, the results are presented as effective $E^\mathrm{eff}_\mathrm{KER}$ distributions. Details are provided in Appendix~\ref{appsec:img}. Here, we focus on the interpretation of the measurements.

The internal states for the DR products are identified using conservation of energy. For ArH$^+$, $E_\mathrm{KER}$ can be predicted from known atomic and molecular properties as
\begin{equation}\label{eq:ker}
    E_\mathrm{KER} = E_\mathrm{I}(\mathrm{H}) - D_0(\mathrm{ArH}^+) - E(\mathrm{Ar}) - E(\mathrm{H}) + E(\mathrm{ArH}^+) + E_\mathrm{c},
\end{equation}
where $E_\mathrm{I}(\mathrm{H})$ is the ionization energy of H; $D_0(\mathrm{ArH}^+)$ is the dissociation energy of ground state ArH$^+$; $E(\mathrm{Ar})$, $E(\mathrm{H})$, and $E(\mathrm{ArH}^+)$ are the respective internal excitation energies of Ar, H, and ArH$^+$, relative to their ground levels; and $E_\mathrm{c}$ is the collision energy. For a nominal $E_\mathrm{c} = 0$~eV, DR of ground state ArH$^+$ X\,$^{1}\Sigma^{+}(v=0, J=0)$ producing ground state Ar$(^1\mathrm{S})$ and H$(n=1)$ results in an expected $E_\mathrm{KER} = 9.7$~eV. For all excited atomic states at a nominal $E_\mathrm{c} = 0$~eV, the calculated $E_\mathrm{KER}$ results in negative values. This indicates that those states can only be reached when additional collision energy or internal energy of ArH$^+$ are added into the system. For ground state ArH$^+$, it is useful to express the threshold for the formation of specific final states as the reaction energy
\begin{equation}\label{eq:re}
    E_\mathrm{r} = D_0(\mathrm{ArH}^+) + E(\mathrm{Ar}) + E(\mathrm{H}) - E_\mathrm{I}(\mathrm{H}).
\end{equation}
Table~\ref{tab:drenergy} lists the four lowest excited product-state combinations relevant for our measurements.

\begin{figure}[ht!]
\includegraphics{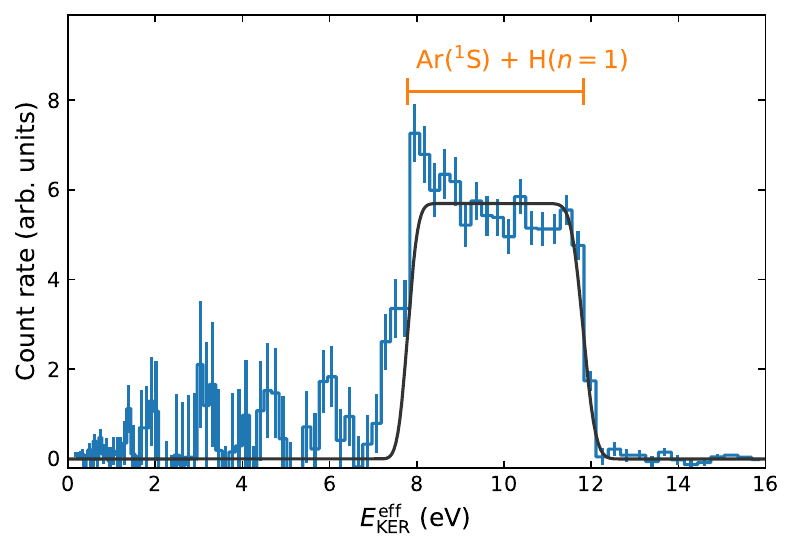}
\caption{Measured $E^\mathrm{eff}_\mathrm{KER}$ distribution for DR of stored ArH$^+$ ions at matched velocities ($E_\mathrm{d} = 0$~eV). The electron-induced signal is plotted in arbitrary units vs $E^\mathrm{eff}_\mathrm{KER}$. The results are shown for events with $\vartheta < 20^\circ$ and $\vartheta > 160^\circ$ as the histogram with one-sigma statistical counting uncertainties. The horizontal orange line with vertical bars is labelled by the DR product states for Ar and H. It shows the range where signal corresponds to $E_\mathrm{KER} = 9.7$~eV and an effective interaction length of $\Delta L = 0.79$~m. The black full line shows the modelled distribution using Eq.~(\ref{eq:distEker}).
\label{fig:3d0eV}}
\end{figure}

The measured $E^\mathrm{eff}_\mathrm{KER}$ distribution for DR at matched velocities ($E_\mathrm{d} = 0$~eV) is plotted in Fig.~\ref{fig:3d0eV}. The data are shown for orientation angles of $\vartheta < 20^\circ$ and $\vartheta > 160^\circ$. These angles represent orientations of the internuclear axis, relative to the bulk electron-ion velocity vector, for which the particle detector can collect all fragments, independent of either $E_\mathrm{KER}$ or the distance from the dissociation location to the detector. Under these conditions, we can compare our data with the empirical model given by Eq.~(\ref{eq:distEker}). By using an effective interaction length of $\Delta L = 0.79$~m, and parameters $\delta$ and $I$ determined by least squares fitting, we can reproduce well the observed data for an $E_\mathrm{KER} = 9.7$~eV expected for Ar($^1$S) + H($n=1$).

\begin{figure}[ht!]
\includegraphics[width=0.5\columnwidth]{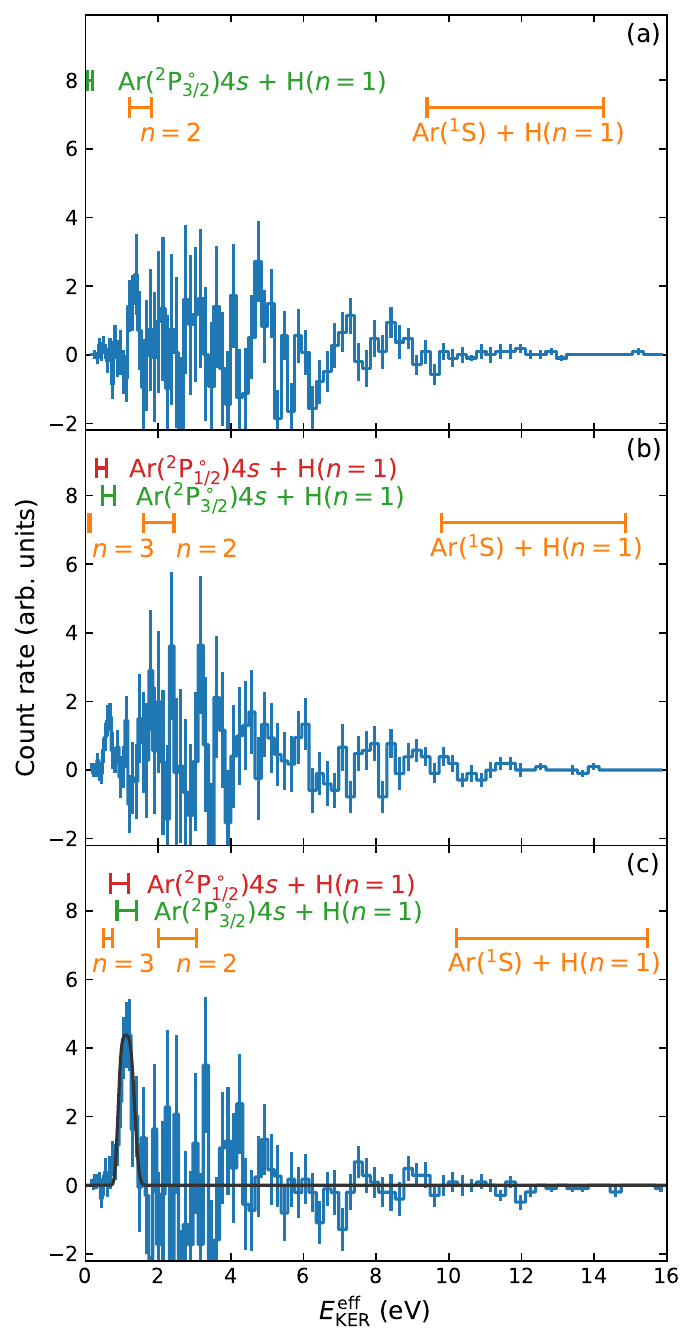}
\caption{Same as Fig.~\ref{fig:3d0eV} but for all values of $\vartheta$ and three values of $E_\mathrm{d}$ above the $1.9$~eV excitation threshold for the production of excited Ar final states. The horizontal lines with vertical bars are labelled by the various DR product states for Ar and H. They show the energy ranges where signal would correspond to the possible product channels and an effective interaction length of $\Delta L = 0.79$~m. Data are shown for (a) $E_\mathrm{d} = 2$~eV, (b) $2.5$~eV, and (c) $3$~eV. The black full line in (c) shows the modelled distribution using Eq.~(\ref{eq:distEker}) for $E_\mathrm{KER} = 1.1$~eV.
\label{fig:3d2to3eV}}
\end{figure}

We also measured $E^\mathrm{eff}_\mathrm{KER}$ distributions for DR at collisions energies between $2$ to $3$~eV, at which several of the excited product states given in Table~\ref{tab:drenergy} are energetically accessible. The data are plotted in Fig.~\ref{fig:3d2to3eV} without restrictions on $\vartheta$ in order to enable the highest sensitivity for weak signals. According to the calculations of Refs.~\cite{abdoulanziz_theor_2018, Djuissi_2022}, Ar states with an electron excited to the $4s$ orbital are expected to be the dominant products for DR at these energies. This has been confirmed in our data for $E_\mathrm{d} = 3$~eV. We also aimed to detect if any of the other energetically accessible states listed in Table~\ref{tab:drenergy} were produced. However, we found no signal within the statistical quality of the data.

\subsection{\label{subsec:kindata} Kinetic Rate Coefficient}

\begin{figure}[ht!]
\includegraphics{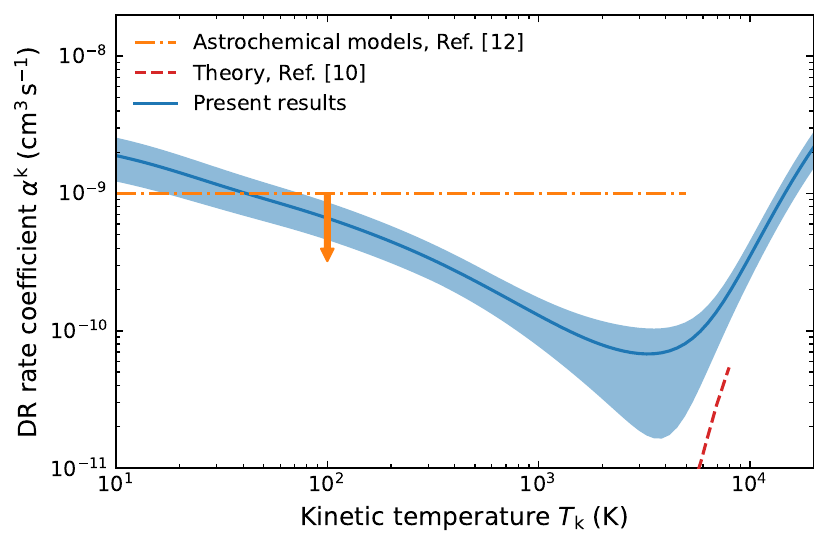}
\caption{Comparison of the kinetic-temperature-dependent DR rate coefficient $\alpha^\mathrm{k}$ from our present experiment to previously published data. Here, the downward pointing arrow indicates that the value of $1\times10^{-9}$~cm$^{3}$\,s$^{-1}$ was taken as an upper limit in Ref.~\cite{schilke_argonium_2014}. The shaded area around the present results corresponds to the total systematic uncertainty of the measurement, mainly due to the absolute scaling of $\alpha^\mathrm{mb}$ and the uncertainty of $T_\perp$. The particular contributions of these uncertainties have been discussed previously \cite{paul_dr_2022}. In the region of the lowest values, the statistical counting uncertainties were also taken into account, as described in the text.
\label{fig:tkin}}
\end{figure}

For use in chemical models, we have generated a kinetic-temperature-dependent rate coefficient $\alpha^\mathrm{k}(T_\mathrm{k})$ from our $\alpha^\mathrm{mb}(E_\mathrm{d})$ results, following the DR cross-section-extraction method from our previous work \cite{novotny_drhcl_2013, paul_dr_2022}. Here, $T_{\mathrm k}$ is the kinetic temperature of the gas, which characterizes the Maxwell-Boltzmann velocity distribution for all particles. We have determined $\alpha^\mathrm{k}$ from $T_\mathrm{k} = 10$ to $20,\!000$~K, as plotted in Fig.~\ref{fig:tkin}. For this temperature range, the possible contributions of DE, as discussed previously, have no impact on our $\alpha^\mathrm{k}$. We provide fitting formulae for the rate coefficient in Appendix~\ref{app:ratecoeff}. The total systematic uncertainty of $\alpha^\mathrm{k}$ at most temperatures is due to the  absolute scaling of $\alpha^\mathrm{mb}$ and the uncertainty of $T_\perp$; however, around the minimum of $\alpha^\mathrm{k}$, the statistical counting uncertainty contributes as well. To determine the lower error band in the cross-section-extraction procedure, we have set to zero all $\alpha^\mathrm{mb}$ values for which the statistical counting uncertainty exceeds the measured value. Also shown in the figure are the experimentally derived upper limit used in astrochemical models \cite{schilke_argonium_2014} and the theoretical DR results of Refs.~\cite{abdoulanziz_theor_2018, Djuissi_2022}.

\section{\label{sec:pec} Theoretical Description}

To theoretically describe the electronic states relevant for DR, it is crucial to have a balanced description of both the neutral Rydberg states converging to the ground state of the ion, as well as the electronic resonant states that cross the ionic potential. These electronic resonance states are neutral Rydberg states with an excited ionic core. The PECs of ArH$^+$ and electronically excited states of ArH have previously been computed in Ref.~\cite{Kirrander_2006} using the multi-reference averaged quadratic coupled cluster method.

\begin{figure}[ht!]
\includegraphics{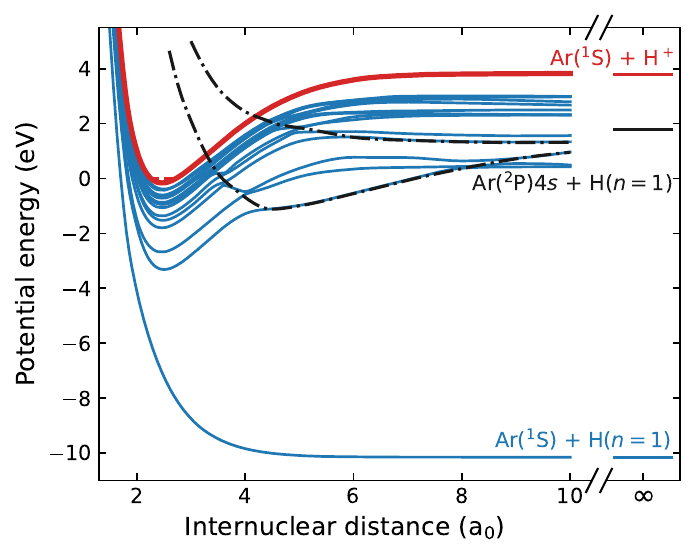}
\caption{Adiabatic PECs of ArH in $^2\Sigma^+$ symmetry (thin blue curves), as well as the PEC of the ground electronic state of ArH$^+$ (thick red curve) and its lowest vibrational level (dashed red line). The dot-dashed black curves display the potentials of quasidiabatic resonant states that contribute to the direct DR. The horizontal full lines show the dissociation limits as given in Fig.~\ref{fig:schempot}.
\label{fig:pec}}
\end{figure}

In Fig.~\ref{fig:pec}, the presently calculated adiabatic PECs of ArH in $^2\Sigma^+$ symmetry are displayed as thin blue curves. This symmetry is important for low energy DR of ArH$^+$. The thick red curve is the ground state PEC of the ion. The energy scale is relative to the $v=0$ vibrational level of the ArH$^+$ ion (denoted by the red dashed line). All PECs were calculated using the multi-reference configuration interaction (MRCI) method \cite{knowles_mrci_1992}. To describe both the Rydberg states below the ground state of the ion as well as the neutral resonance states, we used natural orbitals generated using MRCI calculations on the ionic ground state $X^1\Sigma^+$ as well as the first excited $A^1\Sigma^+$ state of the ion. The $1\sigma,2\sigma,3\sigma,1\pi$ (composed primarily of $1s,2s$ and $2p$ Ar atomic orbitals) were kept doubly occupied, and the reference space of five natural orbitals, consisting of $4\sigma,5\sigma,2\pi$ and $6\sigma$, was used. The basis set was composed of $(12s,5p,2d)$ primitive basis functions contracted to $[6s,5p,2d]$ for argon, while $(11s,6p,4d,1f)$ primitive functions contracted into $[8s,6p,4d,1f]$ was used for hydrogen. Single and double external excitations were included when the natural orbitals were calculated. The natural orbitals were then further expanded with $(7s,4p,3d)$ basis functions centered on argon. The adiabatic PECs of ArH were computed using an MRCI in this expanded orbital space.  Again, the lowest five orbitals were frozen. Full CI was carried out in the reference space of the five natural orbitals described above, and in addition single external excitations out of the reference configurations were included. The calculations were carried out using the MESA program~\cite{mesa}.

The dot-dashed black curves in Fig.~\ref{fig:pec} display ``quasidiabatic" PECs of the resonance states. These were abstracted from the structure calculations by analyzing the CI coefficients of the adiabatic wave functions. The resonant states are Rydberg states converging to electronically excited ionic cores. These cores have the $5\sigma$ orbital or $2\pi$ orbital singly excited. When the potentials of these states are above the ion potential, they should be computed using electron scattering calculations as was done in Refs.~\cite{abdoulanziz_theor_2018, Djuissi_2022}. Here, we only want to illustrate the DR mechanism and use potentials extracted from structure calculations.  

Note that the potentials of the resonant states cross the ion potential at internuclear distances much larger than the equilibrium distance of the ion. The resonance states correlate with asymptotic dissociation limits that are energetically closed at low collision energies. The same applies to electronic resonant states in $^2\Pi$ symmetry not displayed in the figure. At low energies, the only state that is open for dissociation is the electronic ground state. The electron is, therefore, initially captured into a ro-vibrationally excited Rydberg state which subsequently pre-dissociates by non-adiabatic couplings to the ground state as observed here experimentally.

\section{\label{sec:discuss} Discussion}

Our combined merged-beams rate coefficient and 3D imaging measurements at low collision energies ($E_\mathrm{d} \lesssim 0.1$~eV) confirmed the existence of a DR pathway for ArH$^+$ that results in ground state atomic products. This pathway was not detected in previous room temperature measurements, due to experimental limitations \cite{mitchell_diss_2005}. In the absence of resonance states that correlate with the ground state atomic limit and cross the ionic PEC, we conclude that the low energy DR of ArH$^+$ is driven by non-adiabatic interactions where dissociation occurs on the loosely bound PEC of electronic ground state ArH.

Predissociation of bound Rydberg states for ArH has been previously observed in electronic spectra and investigated by theory for selected states (e.g., Refs~\cite{theo_rydberg_1994, baskakov_high_2005} and references therein). Calculations predict a complex dependence of the predissociation lifetime on the electronic, vibrational and rotational states. They also highlight the sensitivity of the calculated predissociation lifetimes to the PECs. Additionally, in low energy DR of ArH$^+$, a range of autoionizing intermediate ArH Rydberg states are involved. The complex interplay between autoionization and predissociation is expected to be reflected in the DR mechanism and in the resonant features observed in the DR cross section.

Our present merged-beams rate coefficient already contains hints at the underlying resonant features in the cross section. They are best highlighted in Fig.~\ref{fig:mbrc} when compared to the featureless $E^{-1}$ cross section model under the assumption of a constant DR probability that is independent of energy \cite{drbook}. The measured data show statistically significant deviations at $25$ and $75$~meV in the form of dips, and at $15$~meV in the form of a peak. Also between $0.1$ and $1$~eV, the experimental average value is significantly smaller than our $E^{-1}$ cross section model. The identification of the corresponding intermediate ArH Rydberg states at these collision energies is beyond the scope of this paper. Our results will be a stringent test for future theoretical cross section calculations down to collision energies of a few meV. For theoretical models incorporating the rotational structure, a comparison can be made by using the experimental rotational level populations, given in Appendix~\ref{appsec:rotmodel}.

For energies $\gtrsim 1$~eV, our measured $\alpha^\mathrm{mb}(E_\mathrm{d})$ closely follows the theoretical results of Refs.~\cite{abdoulanziz_theor_2018, Djuissi_2022}. Minor discrepancies can be seen near the threshold for the first excited Ar products at $\approx 2$~eV, where our values systematically exceed the theoretical ones. This could result from a limitation of the cross section conversion method, inaccuracies of the theoretical model close to threshold, or a contribution of the excited H$(n=2)$ DR products. The latter, even though not correlated with any of the theoretical ArH resonance states, are energetically accessible. We made a dedicated search using our 3D imaging technique between $2$ to $3$~eV for H$(n=2)$ products but found no statistically significant signal. For the present measurements we conclude that a ten times or more greater sensitivity would be required to gain further insights into this matter.

There is an evident discrepancy at low energies between our present results and the theory of Refs.~\cite{abdoulanziz_theor_2018, Djuissi_2022}. This can be explained by the absence of the ground state for ArH as a potential dissociation pathway in the published theoretical model. This state will need to be included in any future model trying to reproduce our present results. The discrepancy at low energies also leads to a considerable difference for $\alpha^\mathrm{k}(T_\mathrm{k})$. The missing cross section contributes even at a few thousand K and the theoretical results fall outside the lower error band of our measurements.

To describe the process theoretically, MQDT calculations are currently being performed; these will be reported in a future study. Compared with the other rare gas diatomic hydrides HeH$^+$ and NeH$^+$, the theoretical description of DR of ArH$^+$ is more challenging. This is not only because there are more electrons in the system, but also because it is crucial to have a balanced description of both the Rydberg states and the doubly excited resonant states. The HeH and NeH systems have no doubly excited states interacting with the Rydberg manifolds situated below the ground state of the ions. Preliminary calculations show that the mixing of orbital angular momentum of the scattered electron, manifesting as off-diagonal elements in the quantum defect matrix $\mu_{ll'}(\epsilon,R)$, is essential to include in the theoretical model. Here, $l$ and $l'$ are the orbital angular momenta of the incoming and scattered electrons, respectively, $\epsilon$ is the scattering energy, and $R$ is the internuclear distance of the target ion. These matrix elements can be extracted from fixed nuclei electron scattering calculations as was done, e.g., in Ref.~\cite{curik_dissociative_2020}, where the quantum defect matrix was obtained from the background phase shifts of the scattering matrix. The electron scattering calculations have to be performed with origin at the center of charge of the molecular ion \cite{curik_dissociative_2020}. We have found that compared to the resonant part of the scattering matrix that describes the contribution from the scattering resonant states, the background phase shifts are much more sensitive to the accuracy of the electron scattering calculation performed. We will return to this matter in our future study.

For astrochemical applications, it is important to note that our $\alpha^\mathrm{k}(T_\mathrm{k})$ is determined for the experimental rotational population, corresponding to a rotational temperature of $\approx 20$~K. Rotational-level-specific results were beyond the possibilities of the measurements, however, for low temperatures $T_\mathrm{k} \lesssim 1000$~K, we can derive an extreme upper limit for the $J = 0$ level. By assuming that only the $J = 0$ level is responsible for the observed data, the upper limit equals the experimental $\alpha^\mathrm{k}(T_\mathrm{k})$ divided by the $J = 0$ population. For high temperatures, this limit is not recommended, since we do not expect a significant rotational-level-dependence for the underlying cross section.

\section{\label{sec:summ} Summary}

Here, we have reported the first quantitative DR measurement for ArH$^+$ in its lowest rotational levels and at collision energies $\lesssim 1$~eV. We have measured its merged-beams rate coefficient over a wide energy range and on an absolute scale. The measurements were further augmented by 3D imaging to determine the internal state of the DR products and gain insight into the recombination pathways. By combining the results, we have demonstrated that the low energy DR of ArH$^+$ is driven by non-adiabatic interactions and its main dissociation pathway is the electronic ground state of ArH.

In the absence of \textit{ab initio} calculations for the low energy DR pathway, we have compared our result to a featureless $E^{-1}$ cross section. This comparison has demonstrated that the underlying cross section consists of resonant features. Within the framework of the indirect DR mechanism, the features are related to the Rydberg states of ArH. Our results can be directly used to benchmark theoretical models at a higher level of electronic complexity than what is typically considered in the literature. We have also compared our results to theoretical calculations for collision energies $\gtrsim 1$~eV where we have found satisfactory agreement. 

Finally, we have derived a kinetic rate coefficient for use in chemical models from our results. For diffuse clouds and similar environments with low internal excitation, we have determined a stringent upper limit for the $J = 0$ state-specific rate coefficient. In denser and hotter environments, like the Crab Nebula, our results represent a reliable constrain on the DR rate coefficient. In Table~\ref{tab:RateOldA}, we have provided an analytical representation for our data.

\begin{acknowledgments}

Financial support by the Max Planck Society is acknowledged. A. K., D. A. N., D. P., D. W. S., and M. G. W. were supported in part by the NASA Astrophysics Research and Analysis program under 80NSSC19K0969. L. W. I. and S. S. akcnowledge financial support by the Deutsche Forschungsgemeinschaft (DFG, grant no. 431145392). We thank J. Forer for fruitful discussions.

\end{acknowledgments}

\appendix

\section{Rotational Level Population Evolution Model} \label{appsec:rotmodel}

In order to predict the rotational populations for the stored ArH$^+$ ions, we have adopted the collisional-radiative model and methodology developed for our previous studies of CH$^+$ and OH$^+$ ions \cite{kalosi_inelastic_2022, kalosi_diss_2023}. Good agreement was found for CH$^+$ between the modeled and measured rotational populations.

The ArH$^+$ from our ion source are electronically, vibrationally, and rotationally excited. The lowest excited electronic states of ArH$^+$ are of repulsive character and the states above those are predissociating \cite{stolyarov_arhp_2005}. Thus, we expect all the ions to be in the X\,$^{1}\Sigma^{+}$ ground electronic state soon after injection. Calculations for vibrational transition probabilities within the ground electronic state find radiative lifetimes of $< 6$~ms \cite{mitchell_diss_2005}. Given the timescale of our experiments, on the order of several tens to hundreds of seconds, we can safely assume that the ions for our measurements were in their ground electronic and vibrational states.

The rotational energies of ArH$^+$ X\,$^{1}\Sigma^{+}(v=0)$ were calculated using the $B_0$ and $D_0$ spectroscopic constants \cite{brown_pure_1988} as
\begin{equation} \label{eq:rotlevel}
    E_{J} = B_0 [J(J+1)] - D_0 [J(J+1)]^2.
\end{equation}
Levels up to $J=49$ were included in the model. We used the theoretical $2.177$~D dipole moment \cite{cheng_dipole_2007} to calculate Einstein coefficients for spontaneous emission for rotational transitions using Equation~(S1) from Ref. \cite{kalosi_inelastic_2022}.

The ambient radiation field in CSR is accounted for by adopting the two-component model of Ref.~\cite{meyer_radiative_2017}. It is the sum of the thermal radiation of the cryogenic chambers and the room-temperature leaks from various openings in CSR. Here, the cryogenic component is estimated to be $T_\mathrm{low} \approx 6$~K, based on the measured chamber temperatures. The room-temperature component is fixed to $300$~K. Previous work has found its fraction relative to the total photon occupation number to be $\varepsilon = (1.0 \pm 0.3)\times10^{-2}$ \cite{kalosi_inelastic_2022}.

Rotational-level-changing collisions between electrons and ions are also present in our experimental setup \cite{kalosi_inelastic_2022}. To include the effects of these collisions, we used the theoretical electron-impact rotational excitation cross sections from Ref.~\cite{hamilton_electronimpact_2016} for $\Delta J = 1$ and $2$ transitions of ArH$^+$. The corresponding de-excitation cross sections were obtained by applying the principle of detailed balance. Rotational-level-changing merged-beams rate coefficients were then calculated for all $E_\mathrm{d}$ corresponding to those used for our DR rate coefficient measurements. For a given transition, the importance of rotational-level-changing collisions can be shown by the ratio of the Einstein $A_{J^{\prime} \rightarrow J^{\prime\prime}}$ rate ($J^{\prime} > J^{\prime\prime}$) to the $J$-level-changing merged-beams rate coefficient $\alpha_{J^{\prime} \rightarrow J^{\prime\prime}}$. This gives the critical electron density
\begin{equation}
    n_\mathrm{c} = A_{J^{\prime} \rightarrow J^{\prime\prime}}/\alpha_{J^{\prime} \rightarrow J^{\prime\prime}}.
\end{equation}
Our DR measurements here were carried out for ions primarily in the $J \leq1$ levels. For the $J^{\prime} = 1 \rightarrow J^{\prime\prime} = 0$ transition, we calculated $A_{1 \rightarrow 0} = 4.3\times10^{-3}$~s$^{-1}$ and a critical electron density of $(0.9 \pm 0.2) \times 10^{3}$~cm$^{-3}$ at matched velocities. This critical density is an order of magnitude smaller than the typical ring-averaged electron density of $(9.4 \pm 0.9) \times 10^{3}$~cm$^{-3}$ for our measurements, meaning that collisions dominate the rotational-level-changing rates in our experiment for the $J\leq1$ levels.

The DR measurement scheme consisted of two ion storage phases. In the first phase, following injection, we applied electron cooling at matched velocities for $42$~s. For the second phase, we measured for predefined sets of $E_\mathrm{d}$ for $408$~s. The $\alpha^\mathrm{mb}$ data shown in Fig.~\ref{fig:mbrc} have been evaluated for storage times $> 42$~s, where the rotational level populations were within $20\%$ of their final values.

\begin{figure}[ht!]
\includegraphics{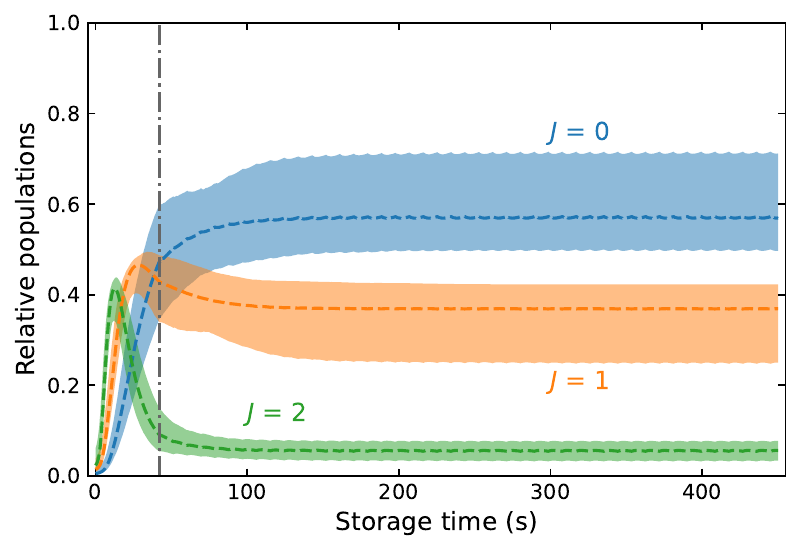}
\caption{Collisional-radiative model of ArH$^+$ rotational-level population evolution as a function of ion-storage time for our DR measurements. The dashed lines are for the mean model with the parameters $k_\mathrm{B}T_\perp = 2$~meV, $T_\mathrm{low} = 6$~K, $\varepsilon = 1.0\times10^{-2}$, and a $3000$~K initial rotational temperature. The accompanying shaded areas indicate the uncertainty of the predictions when varying these parameters, the dipole moment, the electron-impact rotational excitation cross sections, and the initial rotational temperature, all within their estimated uncertainties (see text). The vertical dot-dashed line marks the beginning of the DR data acquisition period.
\label{fig:rotmod}}
\end{figure}

The corresponding modeled relative populations are shown in Fig.~\ref{fig:rotmod} for the levels of interest ($J \leq 2$) as a function of storage time from injection. The initial rotational level populations are given by a Boltzmann distribution within the ground vibrational state. The final population results for levels $J \leq 2$ are insensitive to the initial populations for initial rotational temperatures of $> 300$~K.

The main model uncertainties are due to the experimental parameters $T_\perp$, $T_\mathrm{low}$, and $\varepsilon$. The dipole moment and the electron-impact rotational excitation cross sections for ArH$^+$ can also affect the model predictions. However, the magnitudes of these last two parameters primarily affect the time scale of the rotational population evolution. For the present results, when the populations are close to equilibrium, the uncertainties of these last two parameters do not significantly impact the accuracy of our predictions.

The model uncertainty determination uses the Monte Carlo approach of Ref.~\cite{kalosi_diss_2023}. The dipole moment, electron impact rotational excitation cross sections, and $\varepsilon$ were drawn from normal distributions. For the dipole moment, we estimated an uncertainty of $10\%$, based on the scatter of theoretical results  (e.g., Refs~\cite{cheng_dipole_2007, stolyarov_arhp_2005} and references therein). For the cross sections, the systematic uncertainty at all energies and for all levels was assumed to be $40\%$, based on our CH$^+$ measurements \cite{kalosi_inelastic_2022}. The initial rotational temperature, $T_\mathrm{low}$, and $T_\perp$ were drawn from uniform distributions to include the physical limits on these parameters. For the initial rotational temperature, a lower limit of $300$~K and an upper limit of $5000$~K were chosen. For $T_\mathrm{low}$, the lower and upper limits were set to $4$ and $8$~K, respectively, which have an insignificant impact on the results. For $k_\mathrm{B}T_\perp$, we used the previously determined range from 1 to 3~meV. At each storage time, the uncertainties were evaluated as the 16th and 84th percentiles of the population distributions, which is equivalent to $\pm$~one-sigma for a normal distribution. The shaded areas in Fig.~\ref{fig:rotmod} are given for these percentiles.

Using our model, we estimate the average relative populations of the rotational levels in the measurement time window to be $0.57^{+0.13}_{-0.08}$ for $J=0$, $0.37^{+0.06}_{-0.11}$ for $J=1$, and $0.06^{+0.02}_{-0.03}$ for $J=2$. The uncertainties for the two lowest $J$ levels are anti-correlated and the sum of their populations totals to $\approx 0.94$.

\section{\label{appsec:img} 3D Imaging Analysis}

Imaging the dissociation fragments in our DR measurements enables us to determine the kinetic energy released, $E_\mathrm{KER}$, in the DR process. We use this to identify the internal state of the fragments after dissociation. Unambiguous determination of $E_\mathrm{KER}$ is achieved by combining the measured distance between the fragments in the detector plane $d_\perp$ (assumed to be perpendicular to the ion beam direction) and the difference in the arrival time, $\Delta t$, of the fragments. This method is referred to as 3D imaging (e.g., Ref.~\cite{novotny_thesis_2008}). From $\Delta t$ we determine the distance between the fragments parallel to the beam direction as
\begin{equation}\label{eq:dpar}
    d_\parallel = v_\mathrm{ion} \Delta t,
\end{equation}
where $v_\mathrm{ion}$ is the ion beam velocity in the laboratory frame. From the combination of $d_\perp$ and $d_\parallel$ we can determine the distance between the fragments along the axis of dissociation. This is proportional to the velocity
\begin{equation}\label{eq:vker}
    v_\mathrm{KER} = \sqrt{\frac{2 E_\mathrm{KER}}{M}},
\end{equation}
where $M = m_\mathrm{A} + m_\mathrm{B}$ is the total mass of the heavier $m_\mathrm{A}$ and lighter $m_\mathrm{B}$ fragments.

In our measurements, the velocities of the fragments differ significantly from $v_\mathrm{ion}$ so that we cannot use the approximations of Ref.~\cite{novotny_thesis_2008} to accurately determine $E_\mathrm{KER}$. Here, we included the necessary corrections to the 3D imaging analysis, expanded in the velocity ratio $r_v = v_\mathrm{KER}/v_\mathrm{ion}$.  It was sufficient to implement these corrections to first order in $r_v$. For a detailed derivation of the corresponding equations, see Ref.~\cite{paul_thesis_2021}.

The relationship between $E_\mathrm{KER}$ and the measured distances $d_\perp$ and $d_\parallel$, taking these corrections into account, is given by
\begin{equation}\label{eq:Eker}   
    \sqrt{E_\mathrm{KER}} = \sqrt{E_\mathrm{ion}} \frac{m_\mathrm{A}+m_\mathrm{B}}{\sqrt{m_\mathrm{A}m_\mathrm{B}}}\frac{m_\mathrm{A}m_\mathrm{B}}{\lvert C(s) \rvert}\sqrt{1+ \left(\frac{d_\perp}{d_\parallel} \frac{C(s)}{C(s) - 2m_\mathrm{A}m_\mathrm{B}} \right)^2},
\end{equation}
where
\begin{equation}\label{eq:Cs}   
    C(s) = \frac{s}{d_\parallel}\left(m_\mathrm{A} + m_\mathrm{B} \right)^2 + m_\mathrm{A}^2 - m_\mathrm{B}^2,
\end{equation}
and $s$ is the distance between the dissociation point and the detector. In order to apply this correction, the masses of the fragments have to be determined, which is reflected in the sign of $d_\parallel$. We define $d_\parallel > 0$ when the heavier fragment $m_\mathrm{A}$ arrives first at the detector and negative when $m_\mathrm{B}$ arrives first.

The mass assignment of the imaging data is done in four steps. First, we estimate the average ion beam center projected from the interaction region onto the detector. Second, we assign masses to the impact positions within individual events based on their distance to the projected beam center in the detector plane. Third, we determine center-of-mass positions in the detector plane for all the detected dissociation events and recalculate the projected ion beam center. We repeat the second and third steps until convergence for the projected beam center is achieved. Fourth, to determine which particle arrived first, we make use of the positive correlation between the amplitude of the electrical signals of the MCP (arrival times) and the brightness of phosphor screen spots (impact positions) \cite{urbain_zero_2015}. The signal amplitudes are not clearly correlated with particle mass, but rather establish a connection between the two detection systems. Additional details about the procedure can be found in Ref.~\cite{paul_thesis_2021}.

\begin{figure}[ht!]
\includegraphics{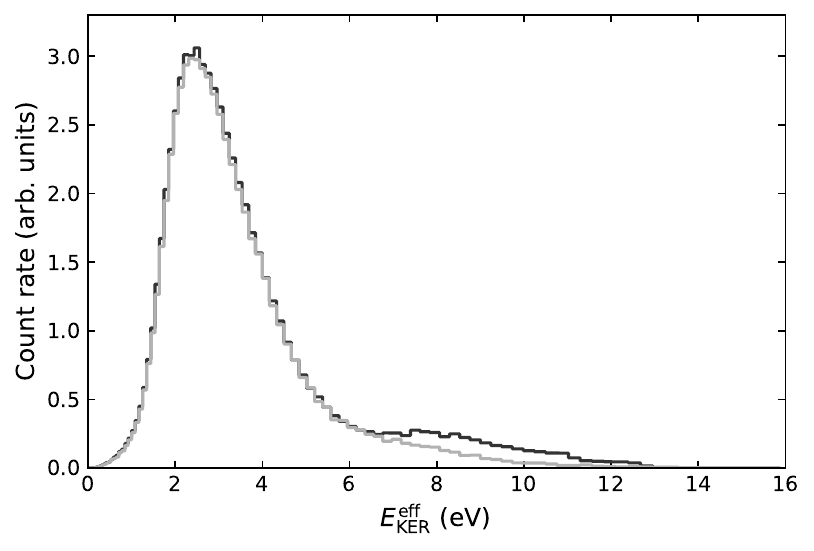}
\caption{Measured $E^\mathrm{eff}_\mathrm{KER}$ distributions in the electrons-off step (gray) and in the measurements step (black) at matched velocities ($E_\mathrm{d} = 0$~eV). The data measured in the electrons-off step is due to residual-gas-induced events. For $E^\mathrm{eff}_\mathrm{KER} \lessapprox 7$~eV, it overlaps with the data recorded in the measurements step. Above $\approx 7$~eV, the difference signal corresponds to DR of ArH$^+$.
\label{fig:3d0eVall}}
\end{figure}

The measured imaging data represents the sum of both DR events and background from residual-gas-induced collisions. The DR events originate within the well-defined overlap region between the electron and ion beams. The mean distance from the overlap region to the particle detector is $L_0 = 3807$~mm. Since we cannot individually determine where each event originates from, we evaluate Eq.~(\ref{eq:Eker}) by approximating $s \approx L_0$, leading to a distribution of effective $E^\mathrm{eff}_\mathrm{KER}$, as shown in Fig.~\ref{fig:3d0eVall}. The background contribution is measured using the $E^\mathrm{eff}_\mathrm{KER}$ histogram for the electrons-off step. This background is subtracted from the measured $E^\mathrm{eff}_\mathrm{KER}$ histogram for the measurement step at $E_\mathrm{d}$. The difference produces the electron-induced signal shown, e.g., in Fig.~\ref{fig:3d0eV}.

The $E^\mathrm{eff}_\mathrm{KER}$ histogram for DR events for a single DR channel with a given $E_\mathrm{KER}$, such as Fig.~\ref{fig:3d0eV}, can be modelled empirically by the combination of a uniform distribution that takes into account the central part of the overlap region where the collision energies are close to $E_\mathrm{d}$ and a Gaussian smoothing that represents the transition into and out of this central region. The uniform distribution is parameterized by the length $\Delta{L}$ of the overlap region. The Gaussian distribution is parameterized by the smoothing parameter $\delta$. This is to account for how in the transition into and out of the central region, the DR signal typically drops rapidly due to the decrease of the DR cross section with increasing collision energy. The resulting weighted probability density function is given as
\begin{eqnarray}\label{eq:distEker}   
    P\left(\sqrt{E^\mathrm{eff}_\mathrm{KER}}\right) = \frac{I L_0}{2\Delta{L}\sqrt{E_\mathrm{KER}}}  &\Biggl[& \erf{ \left(\frac{\sqrt{E^\mathrm{eff}_\mathrm{KER}} - \sqrt{E_\mathrm{KER}} + \sqrt{E_\mathrm{KER}} \Delta{L}/\left(2 L_0\right)}{\sqrt{2}\delta}\right)}  \\  &-& \erf{ \left(\frac{\sqrt{E^\mathrm{eff}_\mathrm{KER}} - \sqrt{E_\mathrm{KER}} - \sqrt{E_\mathrm{KER}} \Delta{L}/\left(2 L_0\right)}{\sqrt{2}\delta}\right)} \Biggr], \nonumber
\end{eqnarray}
where $I$ is a weighting factor so that the integral of the distribution function is equal to that of the measured histogram.

Due to the large mass difference between Ar and H, the dissociation of ArH$^+$ can result in $d_\perp$ larger than the imaging detector radius of 60~mm, starting at $E_\mathrm{KER} \approx 1.5$~eV. Since $d_\perp$ is also a function of dissociation location and distance $s$, the finite size of the detector introduces a complex geometrical cut-off on the measured $E^\mathrm{eff}_\mathrm{KER}$ histogram. To be able to use the empirical model given by Eq.~(\ref{eq:distEker}), we select a subset of the imaging data that is unaffected by the finite detector size. As the selection criterion, we evaluate the polar angle $\vartheta$ between the beam direction and the internuclear axis. This angle is defined as
\begin{equation}\label{eq:vartheta}
    \vartheta = \begin{cases}
        \arctan \left(\frac{d_\perp}{d_\parallel} \frac{C(s)}{C(s) - 2m_\mathrm{A}m_\mathrm{B}} \right) & d_\parallel > 0 \\
        \pi + \arctan \left(\frac{d_\perp}{d_\parallel} \frac{C(s)}{C(s) - 2m_\mathrm{A}m_\mathrm{B}} \right) & \text{otherwise,}
    \end{cases}
\end{equation}
with $s \approx L_0$. By using $\vartheta$ as the selection criterion, we detect events independent of their origin $s$. In particular, we applied this selection criterion to the data plotted in Fig.~\ref{fig:3d0eV}.

The detector set up used in the present measurements has additionally limitations for ArH$^+$ dissociation events with $E_\mathrm{KER} \lesssim 1$~eV. The electronic system is not able to distinguish particle hits with time differences $\Delta t \lesssim 10$~ns. In particular, for $E_\mathrm{KER} \lesssim 0.1$~eV, no 3D imaging data can be collected. The limitations of the electronic readout and detector size translate into a 3D-imaging efficiency as a function of $E_\mathrm{KER}$. It reaches its maximum when $d_\perp$ is close to the detector size, for ArH$^+$ at $E_\mathrm{KER} \approx 1.5$~eV. Above this energy, the efficiency gradually decreases as $E_\mathrm{KER}$ increases. As a result, the data shown in Fig.~\ref{fig:3d2to3eV} only provide a qualitative insight into the branching ratios of possible DR channels.

\section{Beam Purity Analysis} \label{appsec:contam}

The purity of the stored ArH$^+$ ion beam was essential in order to confidently interpret the observed low-energy electron-induced count rate as being due to DR of ArH$^+$. The DR rate coefficient $\alpha^\mathrm{mb}$ of ArH$^+$ was expected to be on the order of $1\times10^{-9}$~cm$^{3}$\,s$^{-1}$ or lower. Any other ions with a fractional contamination of $\gtrsim 1\times10^{-3}$ and an $\alpha^\mathrm{mb} \gtrsim 1\times10^{-6}$~cm$^{3}$\,s$^{-1}$, typical for DR of most molecular ions, would produce a comparable or larger electron-induced signal. However, such a large DR rate coefficient, combined with the ion storage time, also results in the efficient destruction of the contaminants compared to ArH$^+$ in our study.

The initial purity of the stored ArH$^+$ ion beam was established prior to the DR measurements. The ions extracted from the ECR ion source underwent mass selection through a series of dipole magnets prior to injection into CSR. These were sufficient to separate species differing by $\sim 1$~u from ArH$^+$. The stored ion beam was further mass analyzed by operating CSR in the so-called isochronous mode, i.e., as a time-of-flight mass spectrometer with a mass resolution of $\sim 10^{-5}$~u \cite{grieser_iso_2022}. The analysis confirmed the presence of C$_3$H$_5^+$ corresponding to $\sim 1\times10^{-3}$ of the total injected ion number and an even smaller fraction of CH$_3$CN$^+$.

As a follow up on the ion beam composition, we investigated the electrons-on signal at matched velocities as a function of storage time from injection up to $200$~s. During the first $42$~s, the electron beam was tuned to matched velocities and kept on at all times. The electrons-on signal contained contributions from DR, residual-gas-induced collisions, and the detector dark rate, as explained in Subsec.~\ref{subsec:mbexp}. Both the count rate and imaging data revealed a strong time-dependent component in the measured electrons-on signal that we associate with the selective depletion of ion beam contaminants that have large DR rate coefficients.

\begin{figure}[ht!]
\includegraphics{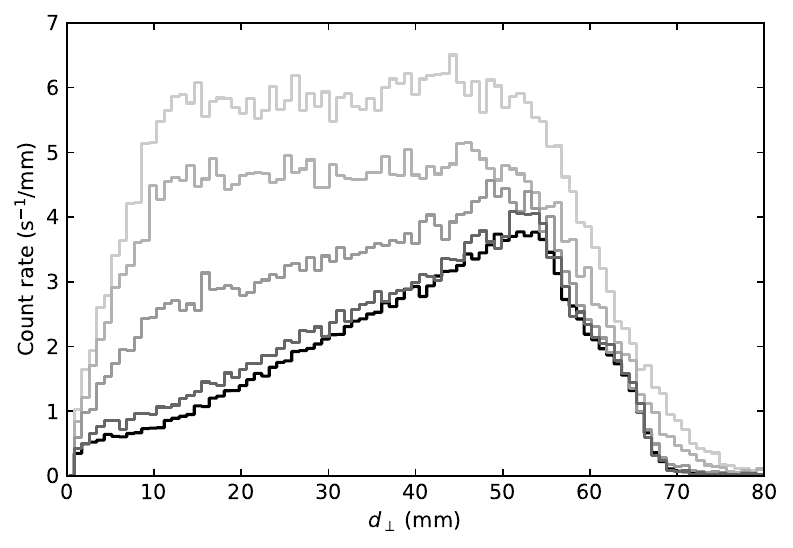}
\caption{Storage-time-dependent $d_\perp$ distributions for electrons on at matched electron--ion velocities. The selected histograms ranging from light to dark gray are for storage times 0 to 5, 5 to 10, 15 to 20, 30 to 35, and 50 to 100~s, respectively. The darkest gray histogram is representative for the signal due to break up of ArH$^+$, present at all storage times. The data at earlier storage times are the sum of the ArH$^+$ and any contaminant ion signal.
\label{fig:2dcontam}}
\end{figure}

The imaging data was used to verify that the time dependent component of the electrons-on signal does not originate from DR of ArH$^+$. We analyzed events that resulted in the detection of two particles. Representative histograms of $d_\perp$ for several storage time windows are shown in Fig.~\ref{fig:2dcontam}. The plotted histograms show a decay of the measured count rate in each bin toward the shape characteristic for all storage times above $\approx 50$~s. This part, present at all storage times, has been identified as the break up of ArH$^+$ due to DR and residual-gas-induced collisions. As indicated by 3D imaging, the break up of ArH$^+$ can lead up to $d_\perp \approx 150$~mm, however, the finite size of the imaging detector introduces a complex cut off for $d_\perp > 50$~mm. The additional signal at storage times $< 50$~s, which we attribute to DR of contaminant ions, is most pronounced at smaller values of $d_\perp$. Residual-gas-induced collisions for contaminant ions are assumed to be negligible, as discussed below. For ArH$^+$ to produce the observed imaging signal at small $d_\perp$, several breakup channels with $E_\mathrm{KER}$ ranging from $0.1$ to $1$~eV would need to be involved. This would imply electronically or vibrationally excited states with $E(\mathrm{ArH}^+) > 0.5$~eV that decay on timescales of several seconds. According to our collisional-radiative model, there are no such states existing for ArH$^+$, thereby ruling out this scenario.

\begin{figure}[ht!]
\includegraphics{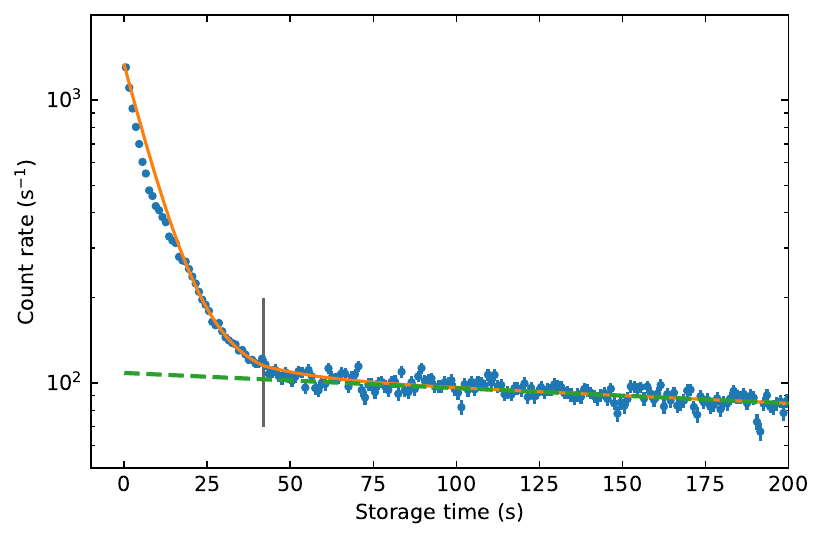}
\caption{Electron-induced count rate as a function of storage time. The measured data are plotted as the symbols with one-sigma statistical error bars. The effective two-component model described in the text is plotted as the orange full line. The model was fitted to the data from $17$ to $200$~s and extrapolated to earlier storage times. The ArH$^+$ component is plotted as the green dashed line. The gray vertical line highlights $42$~s.
\label{fig:contam}}
\end{figure}

The storage-time dependent count rate data further enabled us to estimate $\alpha^\mathrm{mb}$ at matched velocities for the contaminant ions. To be able to separate the DR signal at all storage times, we needed to estimate the non-electron-induced contributions that were monitored only for storage times $> 42$~s in the electrons-off step of our measurement scheme. To account for ion losses, we modelled the storage time dependence of the residual-gas-induced count rate with a single-term exponential function from $42$ to $200$~s and extrapolated it to earlier storage times. This approach assumes that the residual-gas-induced count rate is proportional to the number of stored ions and that it is dominated by ArH$^+$. Even if residual-gas-induced collisions were an order of magnitude more efficient for the contaminant ions, the extrapolation is well justified by the small fraction ($\sim 1\times10^{-3}$) of contaminant ions injected into the storage ring. The extracted electron-induced count rate due to DR of ArH$^+$ and contaminant ions as a function of storage time is shown in Fig.~\ref{fig:contam}.

To quantify the observed storage-time dependence of the electron-induced signal we constructed a model, consisting of the sum of two single-term exponential functions. The slow decaying part represented ArH$^+$ with the same ion loss rate as determined from the residual-gas-induced count rate. The fast decaying part represented the contaminant ions with a fitted DR depletion rate. Additional loss processes for the contaminant ions, e.g., due to ion storage effects, are expected to contribute on the same order of magnitude as observed for ArH$^+$ and are negligible in our analysis. The DR depletion rate of a given species is given by the species-specific $\alpha^\mathrm{mb}$ multiplied with the ring-averaged electron density. In the model, we also took into account the change in effective electron density when additional electrons-off steps were present in the measurement after $42$~s of ion storage. The free parameters of the model were determined from a least squares fit to the measured data for storage times from $17$ to $200$~s, i.e., where the simple model best represents the data. At earlier storage times, the model only qualitatively reproduces the data. This is most likely due to a combinations of the following effects. After injection, the ion beam underwent phase-space cooling that increased the overlap between the ion and electron beams, leading to a change in the effective electron density. Additionally, both ArH$^+$ and contaminant ions are expected to undergo internal cooling that could both increase or decrease their respective $\alpha^\mathrm{mb}$ as a function of storage time. A possibly larger ion loss rate after injection due to ion storage effects when compared to that of a phase-space cooled beam would be underestimated by the extrapolation method.

From our model, we have determined a fitted depletion rate for the contaminant ions to estimate $\alpha^\mathrm{mb} \approx 1\times10^{-5}$~cm$^{3}$s$^{-1}$. This value multiplied by the independently determined fraction of contaminant ions at injection ($\sim 1\times10^{-3}$) indicates an order of magnitude larger electron-induced count rate for the contaminant ions compared with that for ArH$^+$, in agreement with the observed count rate data at the earliest storage times. To summarize, the magnitude and the storage time dependence of the fast decaying electron-induced signal relative to ArH$^+$ indicate the presence of contaminant ions that undergo DR with $\alpha^\mathrm{mb} \approx 1\times10^{-5}$~cm$^{3}$s$^{-1}$ at matched velocities. As a result of this high DR rate coefficient, the stored contaminant ions were rapidly depleted from the ion beam, thereby producing an essentially pure stored ArH$^+$ ion beam.

\section{Kinetic Temperature DR Rate Coefficient} \label{app:ratecoeff}

Here, we provide an analytic representation for the ArH$^+$ DR kinetic temperature rate coefficient $\alpha^\mathrm{k}$ and its one-sigma error band shown in Fig.~\ref{fig:tkin}. We follow Ref.~\cite{novotny_drhcl_2013} and fit $\alpha^\mathrm{k}$ with a function optimized to account for typical DR features, i.e., broad peaks or dips from resonances. The fit function is given by
\begin{eqnarray} \label{eq:RateFunctionON}
	\alpha^{\mathrm{k}}(T_{\mathrm{k}})[\mathrm{cm}^3\, \mathrm{s}^{-1}]&=&
	A\left(\frac{300}{T_{\mathrm{k}}[\mathrm{K}]}\right)^n\\ 
	&+&T_{\mathrm{k}}[\mathrm{K}]^{-1.5}\sum_{r=1}^{N}c_r\exp\left(-\frac{T_r}{T_{\mathrm{k}}[\mathrm{K}] }\right), \nonumber
\end{eqnarray}
and the parameters are listed in Table~\ref{tab:RateOldA}. The maximum relative deviation of the fit is $0.3\%$.

\begin{table*}[b]
\renewcommand*{\arraystretch}{1.4}
\caption{\label{tab:RateOldA}%
Fit parameters for the ArH$^{+}$ kinetic temperature rate coefficient $\alpha^\mathrm{k}$ and its lower and upper error band from Fig.~\ref{fig:tkin}, using Eq.~(\ref{eq:RateFunctionON}).
}
\begin{ruledtabular}
\begin{tabular}{cccc}
\textrm{Parameter}&
\textrm{Rate coefficient}&
\textrm{Lower error limit}&
\textrm{Upper error limit}\\
\colrule
$A$       & ${4.57\times 10^{-10}}$       & ${3.21\times 10^{-10}}$    & ${5.78\times 10^{-10}}$        \\ 
	$n$       & ${0.295}$          & ${0.314}$      & ${0.311}$          \\ 
	$c_1$     & ${6.38\times 10^{-8}}$  & ${2.95\times 10^{-8}}$  & ${8.93\times 10^{-8}}$  \\ 
	$c_2$     & ${-1.93\times 10^{-6}}$  & ${-1.32\times 10^{-6}}$  & ${-2.62\times 10^{-6}}$  \\ 
	$c_3$     & ${-1.04\times 10^{-5}}$  & ${-8.63\times 10^{-6}}$  & ${-1.41\times 10^{-5}}$  \\ 
	$c_4$     & ${-3.05\times 10^{-5}}$  & ${-2.77\times 10^{-5}}$  & ${-3.12\times 10^{-5}}$   \\ 
	$c_5$     & ${-7.50\times 10^{-5}}$  & ${-9.20\times 10^{-5}}$  & ${-7.76\times 10^{-5}}$   \\ 
	$c_6$     & ${1.11\times 10^{-2}}$  & ${1.06\times 10^{-2}}$  & ${1.80\times 10^{-2}}$  \\ 
	$c_7$     & ${1.71\times 10^{-1}}$  & ${1.73\times 10^{-1}}$  & ${2.39\times 10^{-1}}$   \\ 
	$c_8$     & ${8.83\times 10^{-1}}$  & ${7.20\times 10^{-1}}$  & ${1.16\times 10^{0}}$   \\ 
	$T_1$     & ${11.4}$         & ${11.7}$         & ${11.6}$         \\ 
	$T_2$     & ${428}$        & ${421}$        & ${455}$        \\ 
	$T_3$     & ${1180}$        & ${1250}$        & ${1270}$        \\ 
	$T_4$     & ${2930}$       & ${3140}$       & ${3050}$       \\
	$T_5$     & ${7170}$         & ${8180}$         & ${7440}$         \\ 
	$T_6$     & ${41200}$        & ${41700}$        & ${42900}$        \\ 
	$T_7$     & ${77200}$        & ${84600}$        & ${79400}$        \\ 
	$T_8$     & ${139000}$       & ${154000}$       & ${142000}$       \\
\end{tabular}
\end{ruledtabular}
\end{table*}

We also give the Arrhenius--Kooij (AK) representation, as is typically used in astrochemistry, combustion chemistry, and other chemical models and databases. Following the approach of Ref.~\cite{paul_dr_2022}, we provide a set of piecewise-joined fit functions on several temperature intervals, since we cannot model the experimental results with a single AK fit function. This approach introduces discontinuities in the temperature dependence of the analytical rate coefficient between the temperature intervals, which can be avoided by using the representation by Eq.~(\ref{eq:RateFunctionON}). The temperature-interval fit function is given by
\begin{equation} \label{eq:RateFitKIDA}
	\alpha^{\mathrm{k}}(T_{\mathrm{k}})[\mathrm{cm}^3\, \mathrm{s}^{-1}]=A\left(\frac{T_{\mathrm{k}}[\mathrm{K}]}{300}\right)^{\beta} e^{-\frac{\gamma}{T_{\mathrm{k}}[\mathrm{K}]}}.
\end{equation}
The parameters for each temperature interval are listed in Table~\ref{tab:rateKIDA}. The maximum relative deviation of the fit is $10.8\%$. Following the conventions of Ref.~\cite{wakelam_kida_2012}, the relative uncertainty is described by the log-normal factor $F = \exp{\left(\Delta \alpha^\mathrm{k}/\alpha^\mathrm{k}\right)} $. This quantity is fitted on the same temperature intervals as for the AK fits by the fit function 
\begin{equation} \label{eq:RateUncertaintyFitKIDA}
	F(T_{\mathrm{k}})=F_{0}\exp\left(g\left(\frac{1}{T_{\mathrm{k}}[\mathrm{K}]}-\frac{1}{300} \right)\right) .
\end{equation}
Due to the asymmetric error bands of $\alpha^\mathrm{k}$, the log-normal factor is calculated as the average of the upper and lower error bands. The continuity of both the $\alpha^\mathrm{k}$ and error fits are guaranteed on the border of each temperature range.

\begin{table*}[b]
\renewcommand*{\arraystretch}{1.4}
\caption{\label{tab:rateKIDA}%
Fit parameters for the ArH$^{+}$ kinetic temperature rate coefficient $\alpha^\mathrm{k}$ and its relative uncertainty from Fig.~\ref{fig:tkin}, using Eqs~(\ref{eq:RateFitKIDA}) and (\ref{eq:RateUncertaintyFitKIDA}), respectively.
}
\begin{ruledtabular}
\begin{tabular}{ccccc}
\textrm{Parameter}& \multicolumn{4}{c}{\textrm{Temperature range (K)}} \\  &
$10$--$100$&
$100$--$1500$&
$1500$--$7000$&
$7000$--$20000$\\
\colrule
$A$       & ${3.82\times 10^{-10}}$       & ${3.89\times 10^{-10}}$    & ${2.47\times 10^{-13}}$    & ${5.06\times 10^{-13}}$  \\ 
	$\beta$       & ${-0.503}$       & ${-0.861}$    & ${1.758}$  &  ${2.084}$   \\
	$\gamma$       & ${1.027}$       & ${42.2}$    & ${-4680}$   & ${7540}$      \\
\colrule
	$F_0$      & ${1.32}$       & ${1.44}$    & ${0.594}$  & ${1440000}$      \\
	$g$       & ${1.042}$       & ${-11.9}$    & ${-344}$  & ${4260}$      \\ 
\end{tabular}
\end{ruledtabular}
\end{table*}


\bibliography{ArH+Bibliography}

\end{document}